# Automation of finding strong gravitational lenses in the Kilo Degree Survey with U – DenseLens (DenseLens + Segmentation)

Bharath Chowdhary N[1]⋆, Léon V. E. Koopmans[1]⋆, Edwin A. Valentijn,[1] Gijs Verdoes Kleijn,[1] Jelte T. A. de Jong,[1] Nicola Napolitano,[3,4,5] Rui Li,[6,7] Crescenzo Tortora,[2] Valerio Busillo[2] and Yue Dong[8]

[1] *Kapteyn Astronomical Institute, University of Groningen, PO Box 800, NL-9700 AV Groningen, the Netherlands*
[2] *INAF – Osservatorio Astronomico di Capodimonte, Via Moiariello 16, I-80131 Napoli, Italy*
[3] *Department of Physics "E. Pancini", University of Naples Federico II, Via Cintia, 21, 80126 Naples, Italy*
[4] *School of Physics and Astronomy, Sun Yat-sen University, Zhuhai Campus, 2 Daxue Road, Xiangzhou District, Zhuhai 519082, China*
[5] *CSST Science Center for Guangdong-Hong Kong-Macau Great Bay Area, Zhuhai 519082, China*
[6] *School of Astronomy and Space Science, University of Chinese Academy of Sciences, Beijing 100049, China*
[7] *National Astronomical Observatories, Chinese Academy of Sciences, 20A Datun Road, Chaoyang District, Beijing 100012, China*
[8] *Xi'an Jiaotong-Liverpool University, Wuzhong District, Suzhou 215000, China*

**ABSTRACT**
In the context of upcoming large-scale surveys like Euclid, the necessity for the automation of strong lens detection is essential. While existing machine learning pipelines heavily rely on the classification probability (P), this study intends to address the importance of integrating additional metrics, such as Information Content (IC) and the number of pixels above the segmentation threshold ($n_s$), to alleviate the false positive rate in unbalanced data-sets. In this work, we introduce a segmentation algorithm (U-Net) as a supplementary step in the established strong gravitational lens identification pipeline (Denselens), which primarily utilizes $P_{\rm mean}$ and $IC_{\rm mean}$ parameters for the detection and ranking. The results demonstrate that the inclusion of segmentation enables significant reduction of false positives by approximately 25 per cent in the final sample extracted from DenseLens, without compromising the identification of strong lenses. The main objective of this study is to automate the strong lens detection process by integrating these three metrics. To achieve this, a decision tree-based selection process is introduced, applied to the Kilo Degree Survey (KiDS) data. This process involves rank-ordering based on classification scores ($P_{\rm mean}$), filtering based on Information Content ($IC_{\rm mean}$), and segmentation score ($n_s$). Additionally, the study presents 14 newly discovered strong lensing candidates identified by the U-Denselens network using the KiDS DR4 data.

**Key words:** gravitational lensing: strong.

## 1 INTRODUCTION

Strong Gravitational Lensing is caused by the deflection of light from a distant source by a massive foreground object, resulting in multiple (resolved) images, arcs, or rings depending on the nature of the source and alignment. Strong lensing has many applications such as (i) studying the mass distribution of galaxies (Koopmans 2004; Nightingale et al. 2019; Turyshev & Toth 2022), (ii) measuring the Hubble constant $H_0$ using time delays between multiple resolved images (Rhee 1991; Kochanek 2003; Grillo et al. 2018; Treu, Suyu & Marshall 2022; Treu & Shajib 2023), (iii) providing constraints of dark energy (Sarbu, Rusin & Ma 2001; Sereno 2002; Meneghetti et al. 2005; Biesiada 2006; Oguri et al. 2008a) and dark matter (Tortora et al. 2010; Gilman et al. 2019; Nadler et al. 2021) in the Universe, (iv) constraining the slope of inner mass density profile (for e.g. Treu & Koopmans 2002; Zhang 2004; Gavazzi et al. 2007; Koopmans et al. 2009; Zitrin et al. 2012; Spiniello et al. 2015; Li, Shu & Wang 2018; He et al. 2020a; Şengül & Dvorkin 2022), (v) providing constraints on cosmological parameters with lens statistics (Turner, Ostriker & Gott III 1984; Chae et al. 2002, 2004; Mitchell et al. 2005), (vi) act as natural telescopes to study magnified images of distant galaxies (Ellis 2010; Treu & Ellis 2015; Barnacka 2018).

Many strong lenses have been discovered by ground-based and space-based surveys. Space-based surveys such as the Sloan Lens ACS (SLACS) survey (Bolton et al. 2006, 2008; Shu, Bolton & Brownstein 2015; Shu et al. 2017), found up to a few hundred galaxy–galaxy strong lenses using snapshot imaging survey of Hubble Space Telescope (HST). Gavazzi et al. (2008) discovered a double Einstein ring around a gravitational lens using the SLACS survey. Upcoming space-based surveys such as the Nancy Grace Roman Space telescope (previously WFIRST; Wang et al. 2022) is expected to find 17 000 strong lenses (Weiner, Serjeant & Sedgwick 2020). Several thousand strong lenses have also been found in ground-based surveys such as (i) the Kilo Degree Survey (KiDS; de Jong et al. 2013; Kuijken et al. 2019) by Petrillo et al. (2017, 2019a,b), Pearson, Pennock, Clara & Robinson, Tom (2018), Davies, Serjeant & Bromley (2019), Metcalf et al. (2019), Li et al. (2020, 2021), and He et al. (2020b), (ii) the Canada–France–Hawaii Telescope Lensing Survey (CFHTLens; Heymans et al. 2012) by Cabanac et al. (2007), More et al. (2012, 2015), Gavazzi et al. (2014), Sygnet, et al. (2010), Jacobs et al. (2017), and Chan et al. (2015), (iii) The Hyper Suprime-Cam Survey (Miyazaki et al. 2012) by Shu, Yiping et al. (2022), More et al. (2016), Chan et al. (2016), Tanaka et al. (2016), Cañameras et al. (2021), Wong & HSC SSP Strong Lens Working Group (2018), and Jaelani et al. (2020), (iv) VST Optical Imaging of the CDFS and ES1 fields (VOICE; Gentile et al. 2021), (v) Dark Energy Survey (DES; Collaboration 2005) by Rojas et al. (2022), Treu et al. (2018), Diehl et al. (2017), Treu et al. (2018), Anguita et al. (2018), Agnello et al. (2015), Huang et al. (2020), Nord et al. (2015) and Nord et al. (2016, 2020)., and Lemon et al. (2020).

⋆ E-mail: n.bharath.chowdhary@gmail.com (BC); l.v.e.koopmans@rug.nl (LK)

Strong lenses have also been found in other wavebands such as (i) radio imaging based Cosmic Lens All-Sky Survey (CLASS; Myers et al. 2001) by Myers et al. (1999) and Browne et al. (2003), (ii) SDSS Quasar Lens Search (SQLS; Oguri et al. 2006, 2008b) using spectroscopy method by Oguri et al. (2005), Bolton et al. (2006), Belokurov et al. (2009), and Inada et al. (2014), (iii) $u$-band based search with Canada France Imaging Survey (CFIS; Ibata et al. 2017) using multiband Ultraviolet Near Infrared Optical Northern Survey (UNIONS; Savary et al. 2021, 2022). Few hundred strong lenses have been found in sub-mm wavelength with Submillimeter Array (SMA; Negrello et al. 2010), Herschel Multi-tiered Extragalactic Survey (HerMES; Wardlow et al. 2012) and with the Atacama Large Millimeter Array (ALMA; Wootten 2003; Hezaveh et al. 2013).

The number of galaxy-scale strong lens candidates will increase by three orders of magnitudes with upcoming large-scale sky surveys. Around $10^5$ strong lenses are expected to be discovered (e.g. Pawase et al. 2014; Serjeant 2014; Collett 2015) by upcoming large-scale sky surveys such as the Large Synoptic Survey Telescope (LSST; Tyson 2002), Euclid (Laureijs et al. 2010), the Square Kilometer Array (SKA; Dewdney et al. 2009, Koopmans, Browne & Jackson 2004; Quinn et al. 2015), and the Chinese Space Station Telescope (CSST; Zhan 2018). Using human volunteers as classifiers becomes increasingly difficult (next to impossible) with these upcoming surveys. Davies (2022) showed that human classifiers were less successful when compared with the Convolutional Neural Network (hereafter CNN) at classifying strong lenses when subjected to a classification task in a Zooniverse (Simpson, Page & De Roure 2014) project. CNNs have also been greatly preferred after showing promising results in the strong gravitational lens finding challenge (Metcalf et al. 2019).

A Convolutional Neural Network (CNN; Lecun et al. 1998) is an adaptive learning algorithm that learns the features of images utilizing spatial hierarchy through gradient-based backpropagation. Owing to its efficiency, CNNs have been largely preferred over other machine-learning techniques (such as SVM, Random Forests) and extensively used in recent research methodologies (Petrillo et al. 2017; Lanusse et al. 2018; Pearson et al. 2018; Pourrahmani; Nayyeri & Cooray 2018; Schaefer et al. 2018; Davies et al. 2019; Metcalf et al. 2019; 2019a,b; Cañameras et al. 2020; Christ et al. 2020; Li et al. 2020, 2021; Gentile et al. 2021; Rezaei et al. 2022) to find strong lenses. However, due to the highly unbalanced nature of the data-set and the close resemblance of some classes of non-lenses with lens candidates, a large number of false positives in the final sample cannot be avoided. In our previous paper (Nagam et al. 2023), we introduced the DenseNet architecture as a significant step towards mitigating false positives. Building upon this foundation, we advocate for an advanced approach to further diminish false positives.

To achieve a further reduction in false positives, we introduce segmentation techniques alongside Convolutional Neural Networks (CNNs). Segmentation is a technique where select pixels of the image are classified into one or many classes. Some of the popular segmentation architectures include Faster R-CNN (Ren et al. 2015), Mask R-CNN (He et al. 2017), Segnet (Badrinarayanan, Kendall & Cipolla 2017), and U-Net (Ronneberger, Fischer & Brox 2015). Faster R-CNN has been used in the morphology classification of radio sources (Wu et al. 2018), detection of L-Dwarfs (Cao et al. 2023), detection and classification of astronomical targets (Jia, Liu & Sun 2020), detection of supernovae (Wu 2020; Guo et al. 2021). Mask RCNN, a successor of Faster R-CNN, has been used in the morphological segmentation of galaxies (Farias et al. 2020; Gu et al. 2023), to detect, classify and deblend astronomical sources (Burke et al. 2019), to detect and classify sources in radio continuum images (Riggi et al. 2023) and to detect and mask ghosting and scattered-light artifacts from optical survey images (Tanoglidis et al. 2022). Segmentation using U-Net was first proposed by Ronneberger et al. (2015) for medical image segmentation. Since then, it has been widely used in various fields. In radio astronomy U-Net has been used to classify clean signal and RFI signatures (Akeret et al. 2017), automatic recognition of RFI (Long et al. 2019). U-net has also been used to segment spiral arms of disc galaxies (Bekki, K. 2021) and denoizing astronomical images (Vojtekova et al. 2020; Qi et al. 2022).

In the field of strong lensing, U-Net has been used for segmenting dark substructure (subhaloes; Ostdiek, Bryan, Diaz Rivero, Ana & Dvorkin, Cora 2022a), to segment blended galaxy pairs (Boucaud et al. 2019), to measure subhalo mass function (SMF; Ostdiek, Bryan, Diaz Rivero, Ana & Dvorkin, Cora 2022b), to find quadruply imaged quasars (Akhazhanov et al. 2022) and to generate neutrino simulations (Giusarma et al. 2019).

In the case of extensive surveys, such as those being carried out with Euclid, which entail the analysis of millions of candidates, the post-processing results from the Denselens pipeline can still yield thousands of candidates requiring daily vetting. For example, output from Denselens pipeline can still have false positives candidates having features such as arcs, background contamination etc., which can closely resemble strong lensing features. Due to the highly unbalanced nature of the data-sets, where typically one sample out of 1000 samples is a mock lens, the number of false positives getting ended up in final sample can be significant.

Hence, to further reduce these false positives in final sample, we explore a novel idea of using a segmentation algorithm (U-Net) to segment images in to the lensed source pixels of the strong lensing candidates and the 'rest' of the field (other sources in the field and the lens galaxy). Typically, U-Net is favoured over alternative versions of R-CNN due to its lighter model structure (fewer parameters), while maintaining comparable efficiency for semantic segmentation tasks (Widyaningrum et al. 2022). We use U-Nets in addition to the DenseLens (Nagam et al. 2023) network (implemented in our previous paper) to classify strong lenses and to reduce false positives in the final sample.

In Section 2, we describe the KiDS data-sets used for classification and rank-ordering. In Section 3, we describe the methodology to segment source pixels. We explain our results in Section 4 and finally we provide our discussion and main conclusion in Section 5 and Section 6, respectively.

## 2 DATA-SETS

The KiDS is a wide-field optical imaging survey operating with a 268 million pixel square CCD mosaic camera (OmegaCAM; Kuijken et al. 2011) mounted on the VLT-Survey Telescope (VST; Capaccioli & Schipani 2011) at ESO's Paranal observatory in Chile. The KiDS survey is the deepest of the three wide area public imaging surveys ever conducted with best observing conditions. KiDS covers around 1350 square degrees of extragalactic sky in four filters ($u,g,r,i$). The $r$-band images have the optimal seeing condition with a median Point Spread Function (PSF) FWHM of <0.7 arcsec and an exposure time of 1800 s. In this paper, we utilize the data from 904 tiles from KiDS DR4 data release (Kuijken et al. 2019). We have used ∼3.8 million $r$-band cutouts of size 101 × 101 pixels which corresponds to the area of 20 arcseconds × 20 arcseconds.

### 2.1 Data selection

We present a detailed account of the methodology employed in creating KiDS cutouts, outlined below. Our study exclusively utilizes $r$-band images, with $g$, $r$, $i$ images showcased solely for illustrative purposes. Our approach is similar to the methodology introduced by Li et al. (2020) and Petrillo et al. (2019a).

**1. Bright Galaxy (BG) sample:** The objective to create BG sample without any colour cuts is that the colour cuts are arbitrary and the colours of the foreground lens can be contaminated by lensing when the Einstein radius is small (Li et al. 2020). We employ two criteria for selecting KiDS cutouts: (i) Setting the parameter SG2DPHOT to 0 to exclusively target galaxies. SG2DPHOT is a flag generated by the automated tool 2DPHOT (Barbera et al. 2008), offering both integrated and surface photometry for galaxies within an image. (ii) Employing SEXTRACTOR (Bertin & Arnouts 1996), we generate catalogue using $r$-band *mag_auto* with the constraint $r_{auto} \leq 21$. This yields approximately 3.8 million cutouts.

**2. LRG sample:** We focus on selecting Luminous Red Galaxies (LRGs) with redshifts ($z$) less than 0.4 (Petrillo et al. 2019a). This involves isolating areas in ($r$–$i$) and ($g$–$r$) colour diagrams based on the following criteria:

$$
\begin{aligned}
&|c_{\text{perp}}| < 0.2, \\
&r < 14 + c_{\text{par}}/0.3 \\
&\text{where,} \\
&c_{\text{par}} = 0.7(g-r) + 1.2[(r-i) - 0.18], \\
&c_{\text{perp}} = (r-i) - (g-r)/4.0 - 0.18.
\end{aligned} \quad (1)
$$

This selection criterion results in approximately 126 000 LRG cutouts.

## 3 METHODOLOGY

In our prior research (Nagam et al. 2023), we introduced a pioneering approach for the classification and rank-ordering of strong gravitational lenses, called DenseLens. To further enhance accuracy and reduce false positives, we introduce an integrated approach, U-DenseLens, which combines DenseLens with a U-Net segmentation network. The integration of DenseLens and U-Net segmentation aims to refine our model's accuracy in identifying strong lens candidates. In Section 3.1, we briefly introduce DenseLens and in Section 3.2, we delve into the application of U-Net for pixel segmentation within input lens candidate images, providing a detailed methodology for training and classification.

### 3.1 DenseLens

Using DenseLens (Nagam et al. 2023), we demonstrated the application of classification and regression ensemble pipeline for the purpose of classifying and rank-ordering strong lenses. Upon providing an input image to the DenseLens algorithm, four densely connected networks generate classification scores within the range of 0–1. The mean of these scores ($P_{\text{mean}}$) is subsequently computed. We select the candidates with $P_{\text{mean}}$ values above a relatively large designated threshold ($P_{\text{thres}}$). DenseLens also uses a metric called the Information content (IC), which aids in ranking images based on the number of resolution elements in noise-less mock lensed images above a brightness threshold, relative to background noise ($\sigma$). It scales with the ratio of this area ($A_{\text{src},2\sigma}$) in units of the PSF area ($A_{\text{PSF}}$), multiplied by the ratio ($R$) of the Einstein radius ($R_{\text{E}}$) over the effective source radius ($R_{\text{eff}}$) explained in equation (2), preventing a high IC value for lenses with a very large source but a small Einstein radius (which would be hard to identify as lens). The images that pass the $P$-value thresholds are inserted into regression networks which are trained to predict IC values. The resultant mean of the outputs from these regression networks is denoted as $\text{IC}_{\text{mean}}$. The filtered candidates are then systematically rank-ordered based on the computed $\text{IC}_{\text{mean}}$ values. Hereafter, the terms $P_{\text{mean}}$ and $\text{IC}_{\text{mean}}$ are written as P and IC for simplicity.

$$
\text{IC} = \left[\frac{A_{\text{src},2\sigma}}{A_{\text{PSF}}}\right] \times R, \quad (2)
$$

### 3.2 Segmentation

To further reduce false positive rates, we introduce an additional segmentation network at the end of the DenseNet pipeline. This augmented configuration, comprising DenseLens and the U-Net segmentation network, is termed as *U-DenseLens*. This integrated approach aims to further refine the accuracy of our model by leveraging the capabilities of U-Net segmentation in the identification of strong lens candidates. In this study, we employ U-Net for finding source pixels within input lens candidate images. Specifically, for lenses, our training approach designates all source pixels as 1 and other pixels as 0. Conversely, for non-lenses, all pixels are trained with a label of 0. A detailed explanation of this methodology is provided in Appendix A.

The U-Net architecture used in this paper is illustrated in Fig. A2. We employ interpolation techniques to resize the initial 101 × 101 pixel input image to a more suitable 256 × 256 pixel configuration. This resizing step is crucial as the U-Net architecture requires the input image size to be divisible by 32, and the dimensions of the bottom layer should not be excessively small. For instance, opting for a scaled input size of 256 × 256 in the first layer results in a layer size of 16 × 16 in the middle layer. Striking a balance is essential, as selecting a larger input size increases computational complexity, while opting for a lower input size may compromise information flow among the middle layers. The resulting output from the U-Net model is down-sampled again to 101 × 101 pixels, with each pixel exhibiting values in the range of 0–1. A notable modification involves employing a sigmoid activation function (Narayan 1997) for the final layer. This modification ensures that each pixel obtained as output in the final layer possesses values within the range of 0–1. To classify source pixels, we set the segmentation threshold ($S_{\text{thres}}$) to 0.6. We determined this threshold value through extensive experimentation. Intuitively, setting the threshold too low would result in selecting pixels for which the U-Net lacks confidence. Conversely, a high threshold could lead to a multitude of candidates with minimal pixels in the segmentation output. Therefore, we opted for a moderate threshold value, striking a balance. As a third metric we use $n_s$, the total amount of classified source pixels above the segmentation threshold ($S_{thres}$).

**Table 1.** Comparison of segmentation scores ($n_s$) for Lens and Non-Lens samples present in the mock-data.

| | Count | Per cent | | Count | Per cent |
|---|---|---|---|---|---|
| $n_s = 0$ | 1342 | 14 | $n_s = 0$ | 674 | 77 |
| $n_s > 0$ and $n_s < 40$ | 4732 | 49 | $n_s > 0$ & $n_s < 40$ | 201 | 23 |
| $n_s \geq 40$ | 3512 | 37 | $n_s \geq 40$ | 0 | 0 |
| (a) Lens | | | (b) Non-Lens | | |

**Table 2.** Distribution of TP and FP in the BG sample with respect to the mean of human classifier votes $s_m$.

| Condition | Red | Yellow | Green | Total |
|---|---|---|---|---|
| $s_m > 0.5$ (TP) | 19 | 53 | 38 | 110 |
| $s_m \leq 0.5$ (FP) | 287 | 435 | 168 | 890 |
| Total | 306 | 488 | 206 | 1000 |

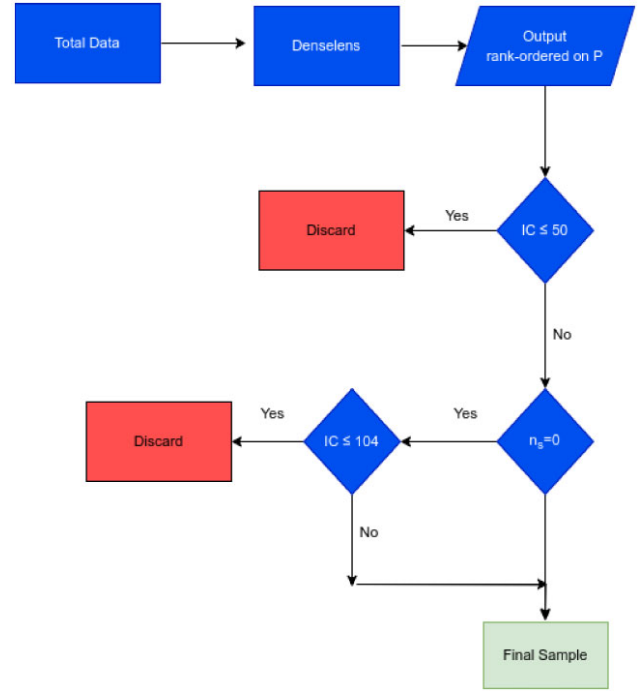

**Figure 1.** Flow diagram for the selection of strong gravitational lenses explained in Section 4.2.

We experimented with mock lenses to investigate diverse threshold values to categorize candidates. This classification included easily identifiable lenses (green), candidates devoid of source pixels (red), and those falling within intermediate classifications (yellow). This leads us to define the following scheme:

$n_s \geq 40$ *green*

$n_s > 0$ and $n_s < 40$ *yellow*

$n_s = 0$ *red*

The values $n_s$, $P$, and IC are utilized to classify and establish a rank order for strong gravitational lenses (using a decision tree; see Section 4.2). By combining multiple metrics, our approach aims to enhance the robustness and accuracy of the classification and ranking process.

## 4 RESULTS

In this section, We systematically generated mock data by combining simulated lensed sources with the LRGs from KiDS (Petrillo et al. 2017) and we have applied our U-Denselens model. We explain the results in Section 4.1. We apply our network and we develop a decision tree based automation technique to BG sample showing the results in Section 4.2. This selection is validated through a voting mechanism with human classifiers, emphasizing the agreement of decision tree results with human classifier votes. Our results demonstrate the efficiency of the proposed approach, offering insights into the reduction of false positives without compromising genuine strong lensing candidates. Additionally, we extended our analysis to the LRG sample in Section 4.3, revealing the versatility and generalizability of our decision tree in optimizing the selection process for diverse data-sets.

### 4.1 Mock data

We generated the mock data as detailed by Nagam et al. (2023), consisting of 10 000 lenses and 10 000 non-lenses. We aim to investigate the impact of segmentation on both classes of mock samples. To achieve this, we deliberately set the threshold parameter ($P_{thres}$) at a low value of 0.3. This choice was made to minimize the extent of filtering applied during the classification prediction ($P$), allowing us to observe the effects of segmentation on the mock sample classes. Following classification with this very low threshold score, we identified 9586 candidates among 10 000 mock lenses as lens candidates and 875 candidates as non-lens candidates. The distribution of candidates is detailed in Table 1(a) for lenses and Table 1 (b) for non-lenses. Notably, Table 1 (b) shows that 77 per cent of non-lens candidates with $P > P_{thres}$ fall into the *Red* category ($n_s = 0$), in other words segmentation does not classify a single pixels as being a lens feature. In contrast, only 14 per cent of the lens candidates share the same categorization. This implies that by excluding candidates in the *Red* category, we can eliminate 77 per cent of false positives, at the cost of only discarding 14 per cent of true lenses. In an unbalanced data set where often only one in a thousand galaxies is a genuine gravitational lens, this drastically reduces the false positive over true positive rate (by about a factor of four). We have also shown the top 25 rank-ordered candidates from the mock data and its respective segmentation maps in Fig. B1 (top) and (bottom) respectively.

### 4.2 Lens candidates from the KiDS bright-Galaxy Sample

#### 4.2.1 General approach

To optimally combine the three metrics ($P$, IC, and $n_s$) and assess their respective impact on the identification, We have devised a decision tree to reduce the false positives in the final sample (see the flow diagram in Fig. 1). Our decision tree for the selection of strong gravitational lenses consists of a number of steps and selection criteria: **(i) Rank-order candidates based on P:** We initiate the selection process by rank-ordering candidates according to their classification scores ($P$).

**(ii) Filtering candidates with IC $\leq$ 50:** The IC quantifies the ranking of images by considering the resolution elements in noiseless mock lensed images above a brightness threshold, with scaling factors based on the ratio of Einstein radius ($R_E$) to the effective source radius ($R_{eff}$) aiming to provide higher IC values for easily recognizable lenses. Candidates characterized by an IC less than or equal to 50 are filtered out in this step. This strategic filtering step is crucial to exclude candidates exhibiting thick blob like structures. Fig. 2 illustrates the 12 highest ranked candidates that fall within this category. Notably, these candidates often present challenges in differentiation due to presence of dense central structures, making them indistinguishable as lenses (except for the first candidate). Despite their high *P* values and elevated ranks, the necessity to eliminate candidates with IC values less than or equal to 50 is apparent. We note that various test have shown the results to be relatively robust against changes in this value. **(iii) Remove candidates with $n_s = 0$ and IC $\leq$ 104:** Subsequent to the IC-based filtration, candidates with no classified source pixels ($n_s = 0$) and IC values less than or equal to 104 are removed. We discuss how we arrived at this value shortly later in the section. This additional step ensures a refined selection process by removing the candidates that are completely rejected by the U-Net segmentation algorithm.

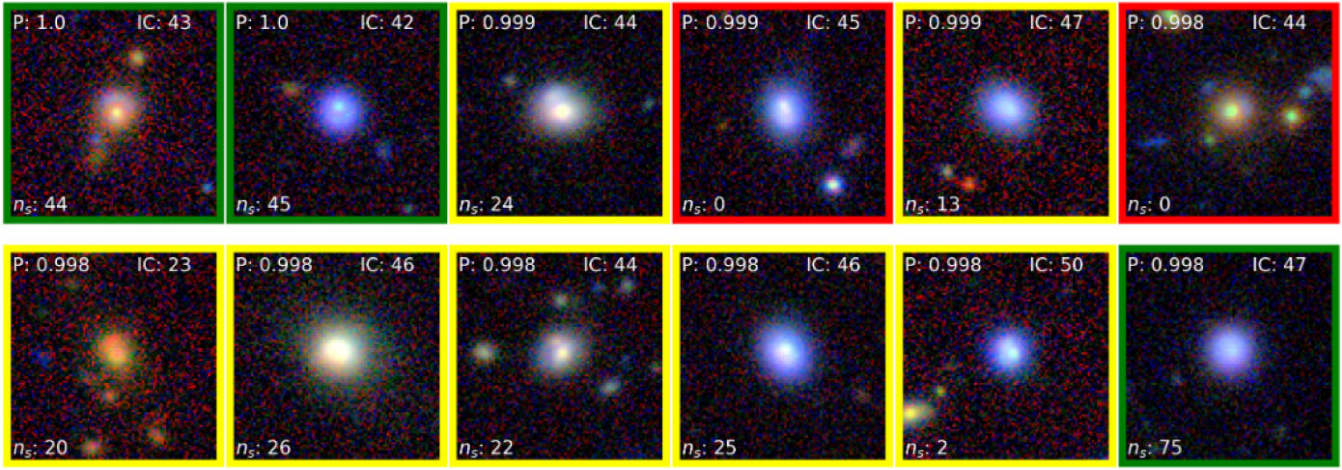

**Figure 2.** High ranked candidates in the KiDS-BG sample (based on P) with IC≤50. Such candidates (including the candidates shown) having IC≤50 were removed before given to the human classifiers for voting. Their details are shown in Table D1.

*4.2.2 Visual inspection and classification*

Throughout this paper, we use the term 'Human classifiers' to collectively refer to the eight authors involved in this study. This group was tasked with evaluating the top thousand candidates selected the BG sample as they had high *P* value ($\sim P \geq$ 0.95), and were rank-ordered solely based on the value of *P*. Each human classifier voted on all candidates, utilizing a set of pre-defined options. Notably, each option is associated with a corresponding score as detailed below. To these four categories, we assign a weight of 1.0, 0.7, 0.3, and 0.0, respectively.

The distribution of candidates in red, yellow, green categories (as defined in Section 3.2) yielded 306, 488, 206 candidates, respectively. Thus we could argue that if we remove these 306 candidates with $n_s = 0$ (red) as false positives, we can potentially reduce the false positives in the final sample by 30 per cent without human culling. However, prior to this removal, it is necessary to maximize the exclusion of only false positives and not genuine candidates. Consequently, we undertook a validation process, automating the identification of strong lenses through a voting mechanism involving human classifiers.

The top 1000 candidates were presented to human classifiers in a randomized, label-free manner to eliminate any potential bias. The voting results, plotted against *P*, are depicted in Fig. 3. Candidates selected by obtaining votes of either 'lens' (a weight of 1) or 'maybe lens' (a weight of 0.7) from four or more human classifiers are marked as triangles and will be referred as *democratically elected samples* throughout this paper. There were 306 candidates with $n_s = 0$. If we define positive samples as the candidates having the mean of human classifier votes ($s_m$) >0.5 and then the reminder as negative samples, then the 306 candidates having $n_s = 0$ (red) split into 19 positive samples and 287 negative samples. Hence, rejecting $n_s = 0$ (red) candidates, enables us to eliminate an additional 287 false positives in final sample. But before we include this selection step, we want to identify how many out of the 19 positive samples can be retained based on an additional selection. In the Fig. 4, we have plotted the IC values against the percentages of TP and FP. The figure describes how many TP (out of 19 positive candidates) and FP (out of 287 negative candidates) passed the IC threshold ranging from lowest IC (50) to the maximum IC of all candidates. We find that at IC = 104, we can make a trade off as it retains approximately 60 per cent of true positives candidates (11 out of 19) and also lowers the false positives to approximately 9 per cent (27 out of 287). We define these candidates as blue candidates which fall in the IC > 104 range, despite $n_s = 0$. The top six of these blue candidates are illustrated in the Fig. 5. The top 6 candidates of the other regimes, namely yellow (0<$n_s$<40), green ($n_s \geq$40), and red ($n_s = 0$) are also shown in the same figure. We have also shown the distribution of TP and FP in the BG sample with respect to the mean of human classifier votes in Table 2.

Instead of arbitrarily defining positive samples as $s_m > 0.5$, we could also define positive samples as *democratically elected samples* that received a majority of human votes (at least 4 out of 8 people) voted as maybe lenses (0.7) or as sure lenses (1). When removing all red candidates in the top 1000 candidates from the BG sample, we recognized a potential loss of 16 democratically selected red candidates. The total count of these democratically elected candidates in Fig. 3 (marked by triangles) is 169, split in to 16 red, 15 blue, 60 green, and 78 yellow candidates. The proposed removal strategy would result in only an $\sim$ 10 per cent loss, specifically sixteen red candidates. Therefore, our validation of segmentation algorithm, based on the votes from human classifiers, underscores that eliminating red candidates could significantly reduce the false positives in the final sample by a quarter, while loosing considerably fewer strong lensing candidates.

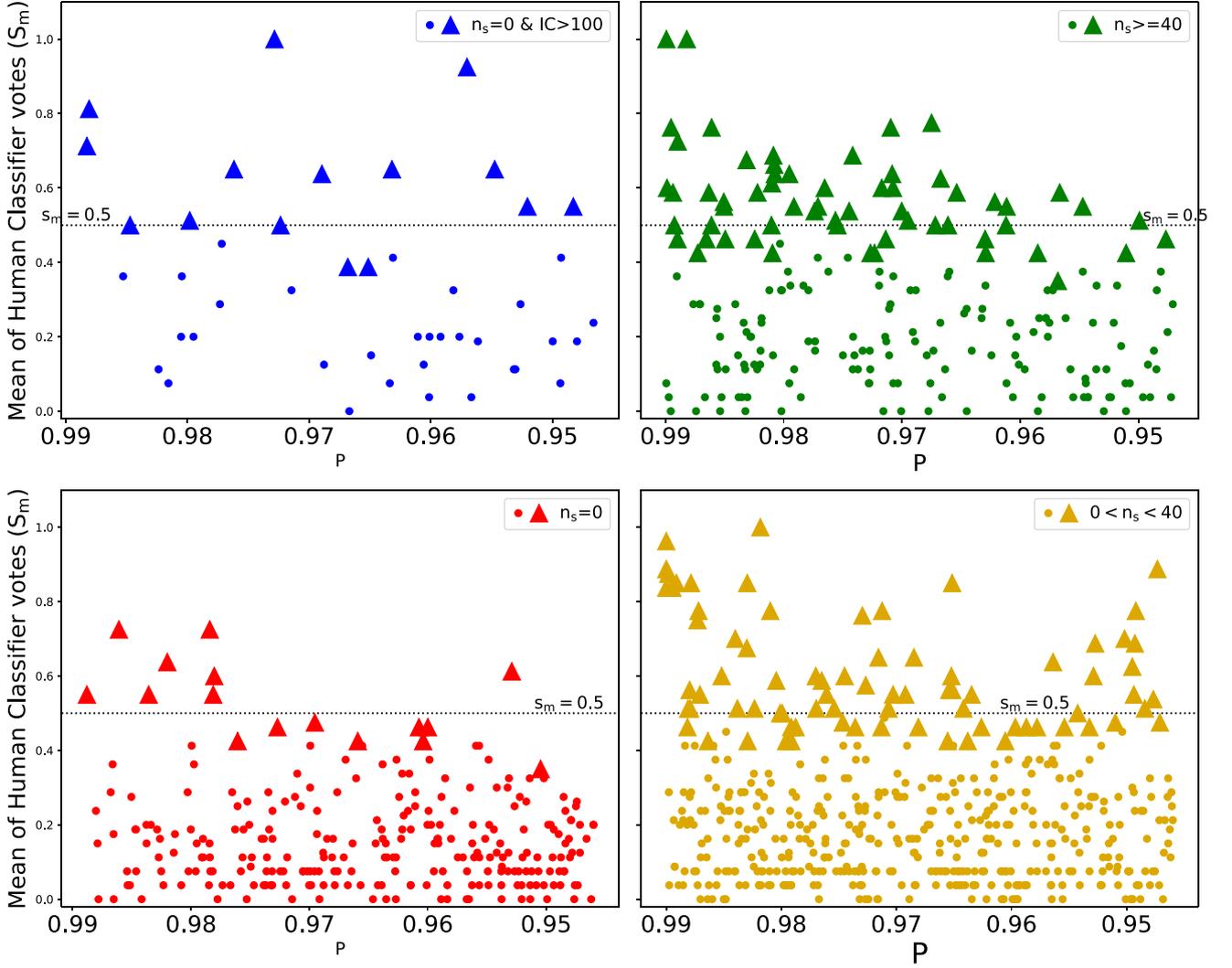

**Figure 3.** Mean of human classifier votes ($s_m$) for all the 1000 candidates from the KiDS-BG sample plotted against $P$. The 1000 candidates are segregated into four plots (blue, green, red, and yellow) based on their observed $n_s$ regime. Candidates which are democratically voted as *May be lens* or *Sure lens* by four or more voters is indicated by triangle shaped marker.

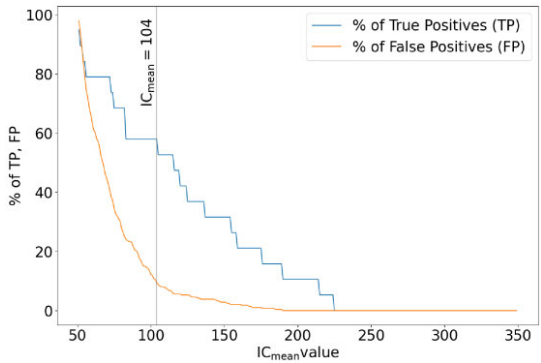

**Figure 4.** Percentage of TP (blue continuous line) and FP (orange continuous line) recovered for given range of IC values. At an IC value of 104, we recover as many as 60 per cent of true positives.

We also have found 14 strong lensing candidates in the BG sample which have not been previously discovered before. These candidates have been voted as lens or may-be lens by four or more human classifiers. The candidates are shown in Fig. 9.

*4.2.3 Random forest analysis*

A critical analysis was performed to discern the primary contributor to decision-making among the three metrics ($P$, IC, $n_s$). The determination of feature importance, is done through a Random Forest model (Breiman 2001), by quantifying the reduction in Gini impurity (Gini 1921) between parent and child nodes, in the decision tree as presented in this work. Detailed computations are provided in Appendix E. A higher reduction in Gini impurity signifies greater importance.

In our analysis, we only use the $P$, IC, and $n_s$ values of the top 1000 candidates (rank-ordered based on the $P$ value) from the BG sample as input features, given that only those were also given to human classifiers. The output values, being the mean of human voting results ($s_m$) rounded to the nearest integer value (0 or 1), were employed for training a random forest comprising 100 000 decision trees. The resulting feature importance are 37.5, 41.0, and 21.5 per cent for P, IC, and ($n_s$), respectively. The feature importance analysis shows that all P and IC have very similar importance, but the ($n_s$) still considerably contributes to the final selection albeit with less weight than P and IC.

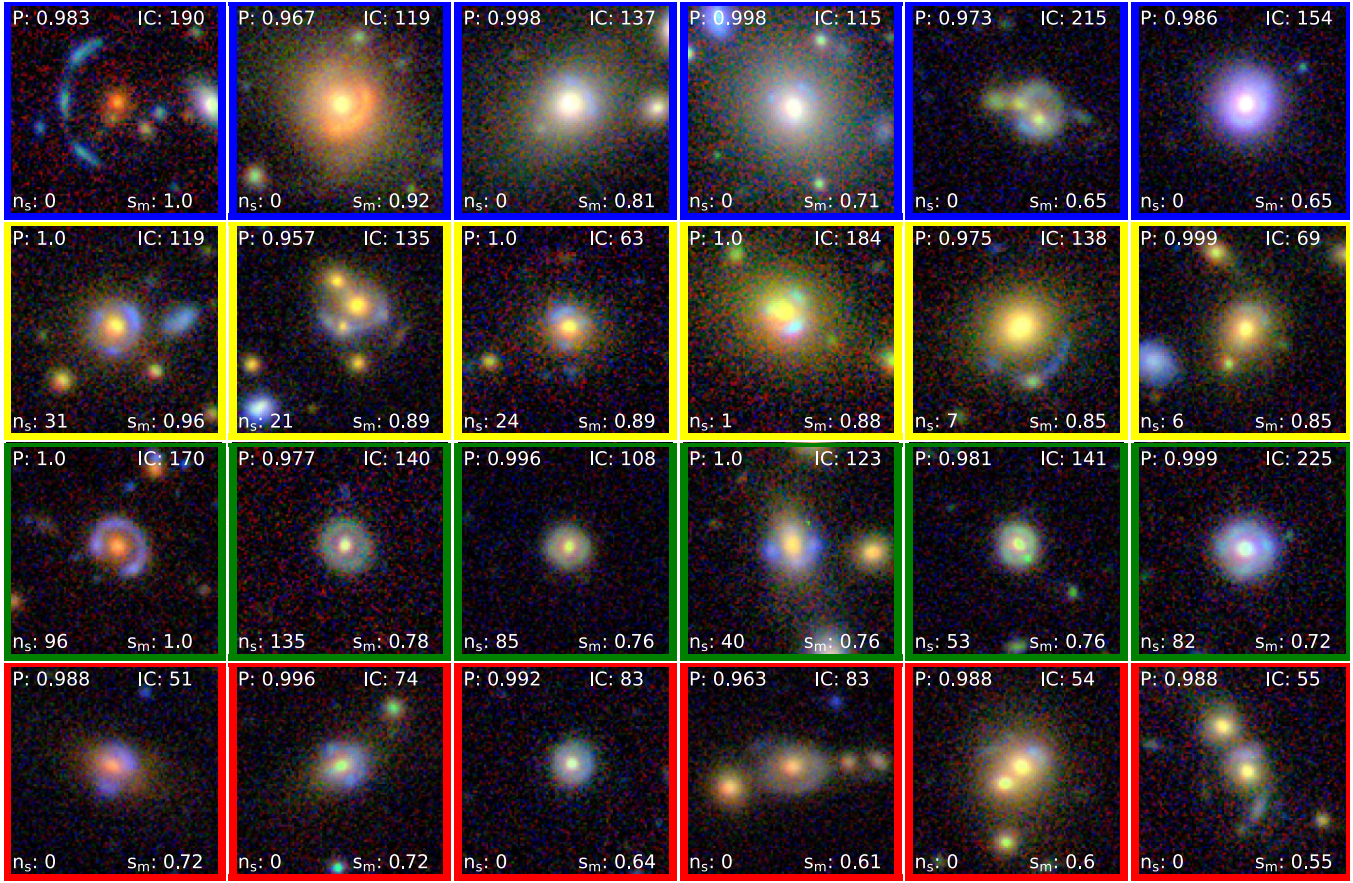

**Figure 5.** The top six candidates based on $s_m$ in each regime from Fig. 3 are shown. Top row: The top six candidates ranked based on $s_m$ with $n_s = 0$ but have IC>104 shown as blue triangles in Fig. 3 (top-left). Second, third, and fourth row: Also shown are the top six candidates ranked based on $s_m$ that have $0 < n_s < 40$, $n_s \geq 40$ and $n_s = 0$, respectively, are shown as yellow, green, and red triangles, respectively in Fig. 3. Their details are shown in Table D2.

### 4.3 LRG sample

We repeated the analysis with the KiDS–LRG sample, but performed the voting with the human classifiers only for the top-200 candidates (sorted based on P) out of ∼126 000 candidates.

We carried out the voting experiment for the LRG sample against the results from the decision tree implemented in Section 4.2. The mean of human classifier votes $s_m$ is plotted against P. This is shown in Fig. 6. There were 9, 116, and 75 candidates in green, yellow, and red regimes, respectively, shown in the Table 3. There are also 8 true positives and 67 false positives in the $n_s = 0$ (red) regime. If we apply the same decision tree (shown in Fig. 1) and by putting a IC threshold of 104 in the final step, we can retain 5 out of 8 TP (60 per cent) at the expense of 14 false positives from the $n_s = 0$ regime. These candidates fall into the 'blue' regime. By removing the 'red' candidates, we remove an additional 56 out of 200 false positives in the final sample reducing approximately a quarter (∼ 25 per cent), by combining the results from P, IC, and $n_s$ in the LRG sample.

Using the similar definition applied to the BG sample, now TP's in LRG samples again can be defined as the candidates selected as *democratically elected samples*. The democratically elected samples are shown as triangle shaped markers in Fig. 6. We see that only three of the red candidates are present in the democratically elected regime. There are only six blues, four green, and twenty one yellow candidates in the human-classifier in the democratically elected regime which are shown in Fig. 7 as triangle shaped markers. If we define true positives as the candidates belonging to democratically elected regime, then there are 34 TPs. By removing red candidates, we will only lose 3 out of 34 TPs (∼ 10 per cent). Thus again we prove the validation of the segmentation algorithm and we can conclude that eliminating 'red' candidates can significantly reduce false positives while loosing only fewer strong lensing candidates. The standard deviation for many candidates are high showing high disagreement among voters for certain candidates. The candidates having highest standard deviation is shown in Fig. 8 and their details in Table D4. This shows that these candidates show features that do not convince all of the voters of it being a genuine lens.

## 5 DISCUSSION

The primary objective of this paper has been to find an algorithm capable of classifying strong lenses without requiring human vet-ting, particularly in very large surveys such as the wide survey carried out with Euclid, which could potentially comprise several hundred thousand strong lensing candidates (Collett 2015). Previous approaches in strong lens classification have predominantly relied on the classification probability (P). Nagam et al. (2023) introduced DenseLens and the concept of combining P and IC to refine candidate selection. After filtering based on *P*, we rank-order candidates using IC. In this work, we introduced segmentation as an additional metric to improve our ability to differentiate final candidates based on whether they contain plausible lensed features.

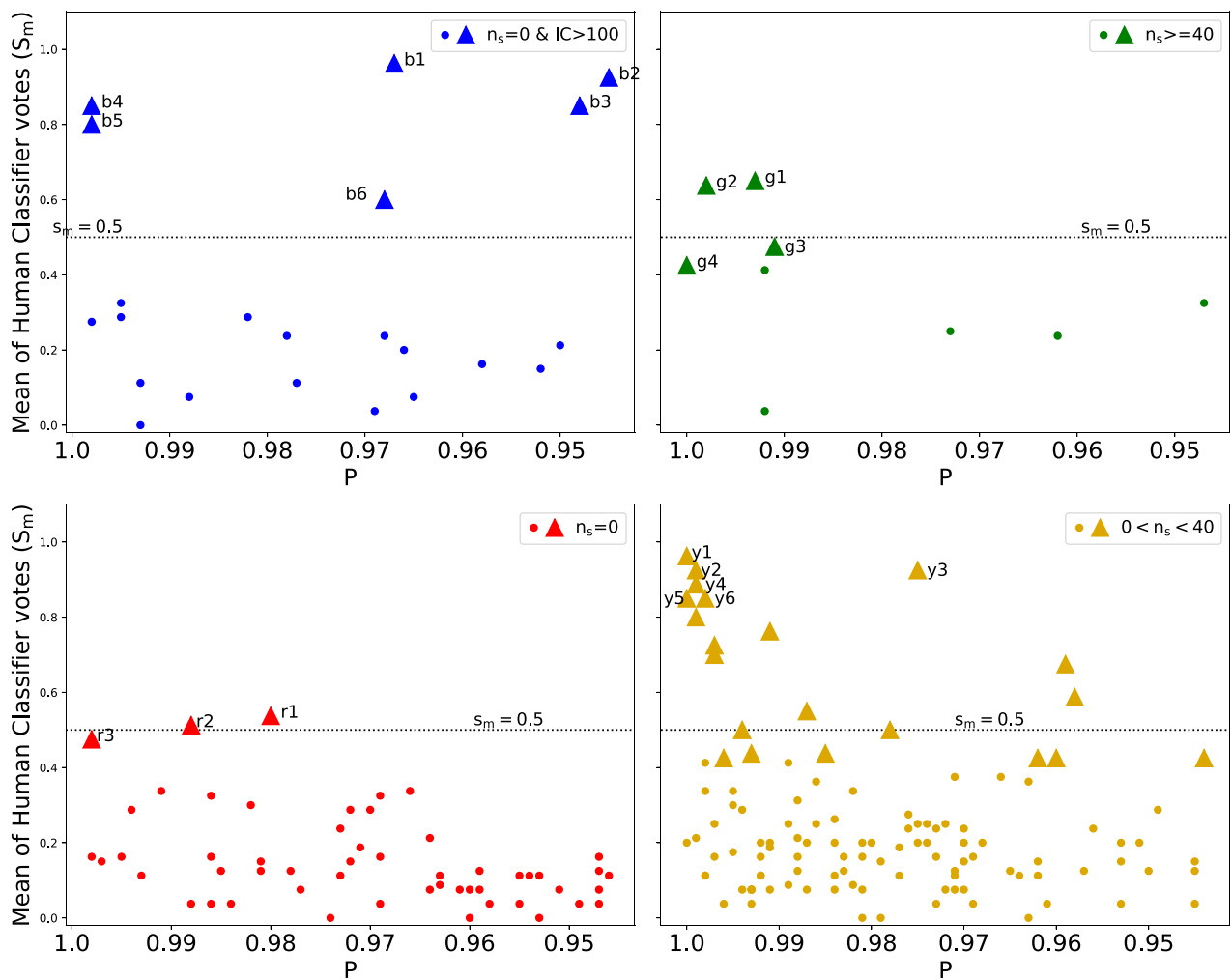

**Figure 6.** Mean of human classifier votes ($s_m$) for all the 200 candidates from KiDS-LRG sample plotted against *P*. The 200 candidates are segregated into four plots (blue, green, red, and yellow) based on their observed $n_s$ regime. Candidates that are democratically voted as *May be lens* or *Sure lens* by four or more voters is indicated by triangle shaped marker.

We propose the idea of considering the number of pixels ($n_s$) above a segmentation threshold ($> 0.6$) as an additional metric to reduce false positives. However, the values of *P*, IC, and $n_s$ are not independent. Hence, we employ a decision tree to combine them. To ensure that the decision tree does not discard too many strong lens candidates during the sample size reduction via segmentation, we set the selection criteria based on voting by human classifiers.

We find that the retention of candidates with $n_s = 0$, when they have a value IC $>104$, ensure the inclusion of highly-plausible lenses without increase the false positive rate significantly. The rationale behind this choice lies in the fact that not all variations of strong lenses are encompassed in the training data-set used for segmentation algorithms, as illustrated that a low value of $n_s$ and a high value of IC are in contradiction since the former suggest there are no lens features (or the features are seen to be from non-lenses) and the latter suggests the opposite. For instance, our segmentation training focused on candidates featuring a single lens in the foreground. However, when examining 'b1' in Fig. 5, it becomes evident that this specific candidate exhibits multiple foreground strong lenses, and larger Einstein radius than that is expected for a single lens galaxy. Consequently, the combination of information content (IC) and the segmentation threshold $n_s$ addresses the mismatch between the training set and real data (leading to the tension between $n_s$ and IC in some cases). This approach recognizes the complexity of strong lensing scenarios, especially when deviations from the training set parameters are encountered, and underscores the need for a comprehensive evaluation that considers both IC values and segmentation thresholds. This tension could be alleviated by training the network on more complex lensing scenarios, but that is outside the scope of this work. Segmentation algorithm, when used with other network architectures involved in automated searches, can be beneficial. However, quantifying the extent of their impact would also be beyond the scope of this paper.

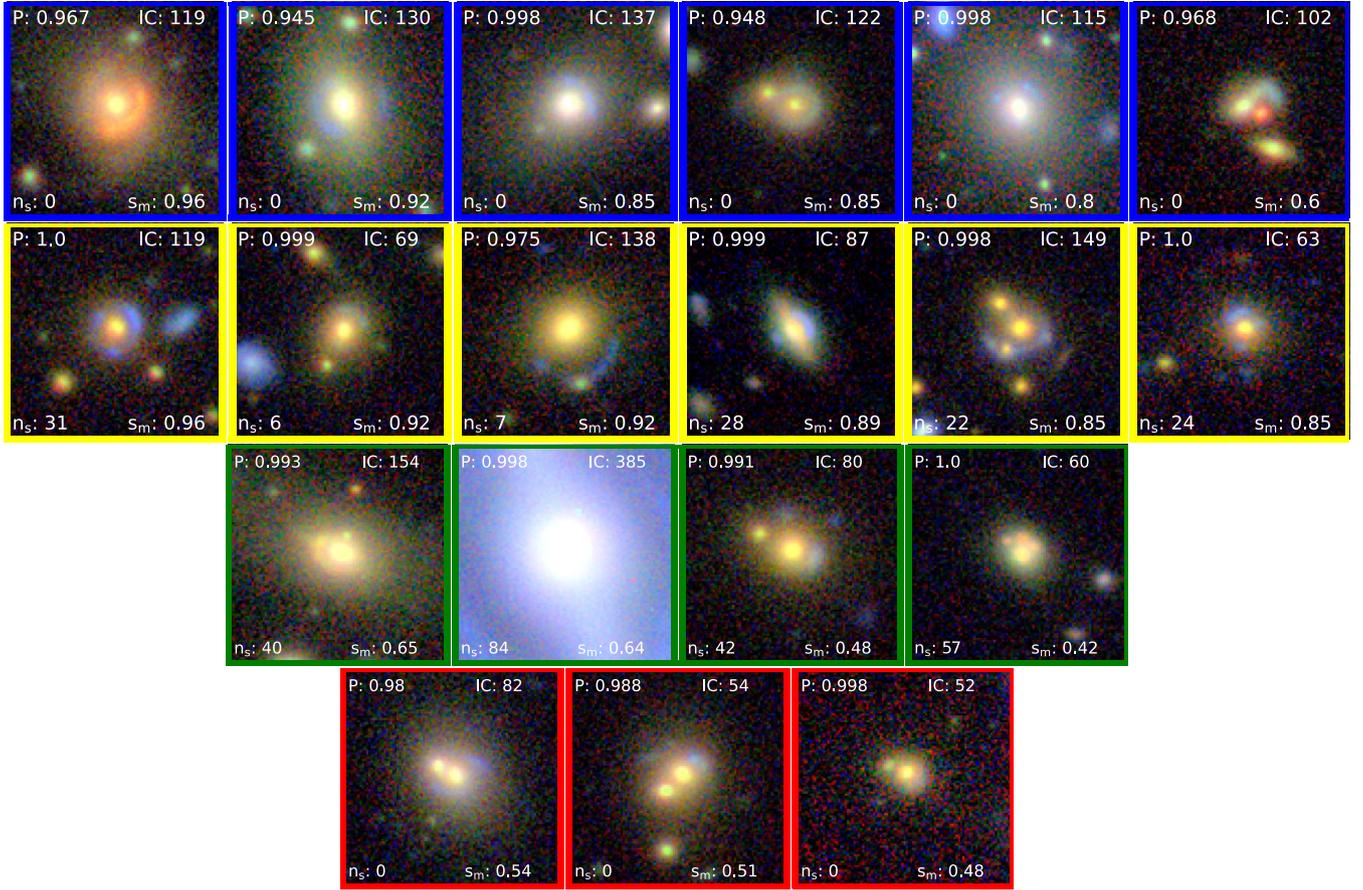

**Figure 7.** LRG candidates tagged as b1–b6 (top row), y1–y6 (2$^{nd}$ row), g1–g4 (3$^{rd}$ row) and r1–r3 (4$^{th}$ row) shown in Fig. 6. Their details are shown in Table D3.

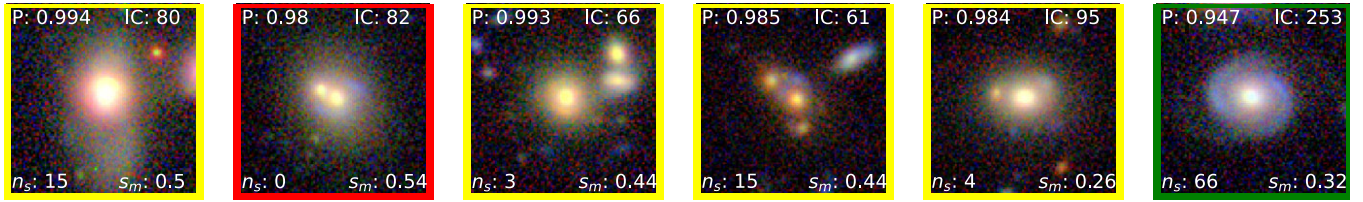

**Figure 8.** LRG candidates with high standard deviation shown in Fig. 6. Their details including standard deviation values $s_{std}$ are shown in Table D4.

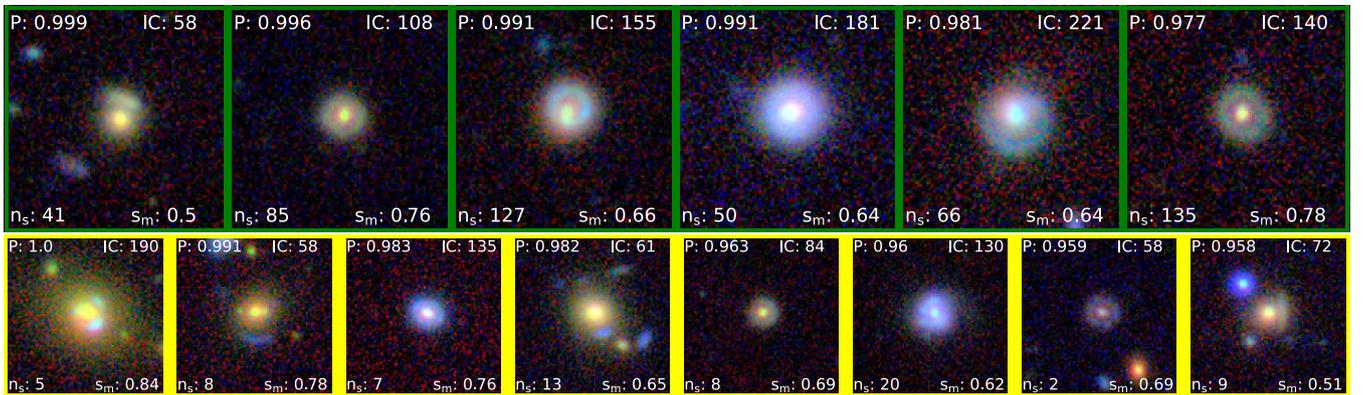

**Figure 9.** Fourteen new strong lensing candidates discovered in BG sample which have been agreed by four or more human classifiers as lens or may be lens. The details are shown in Table D5.

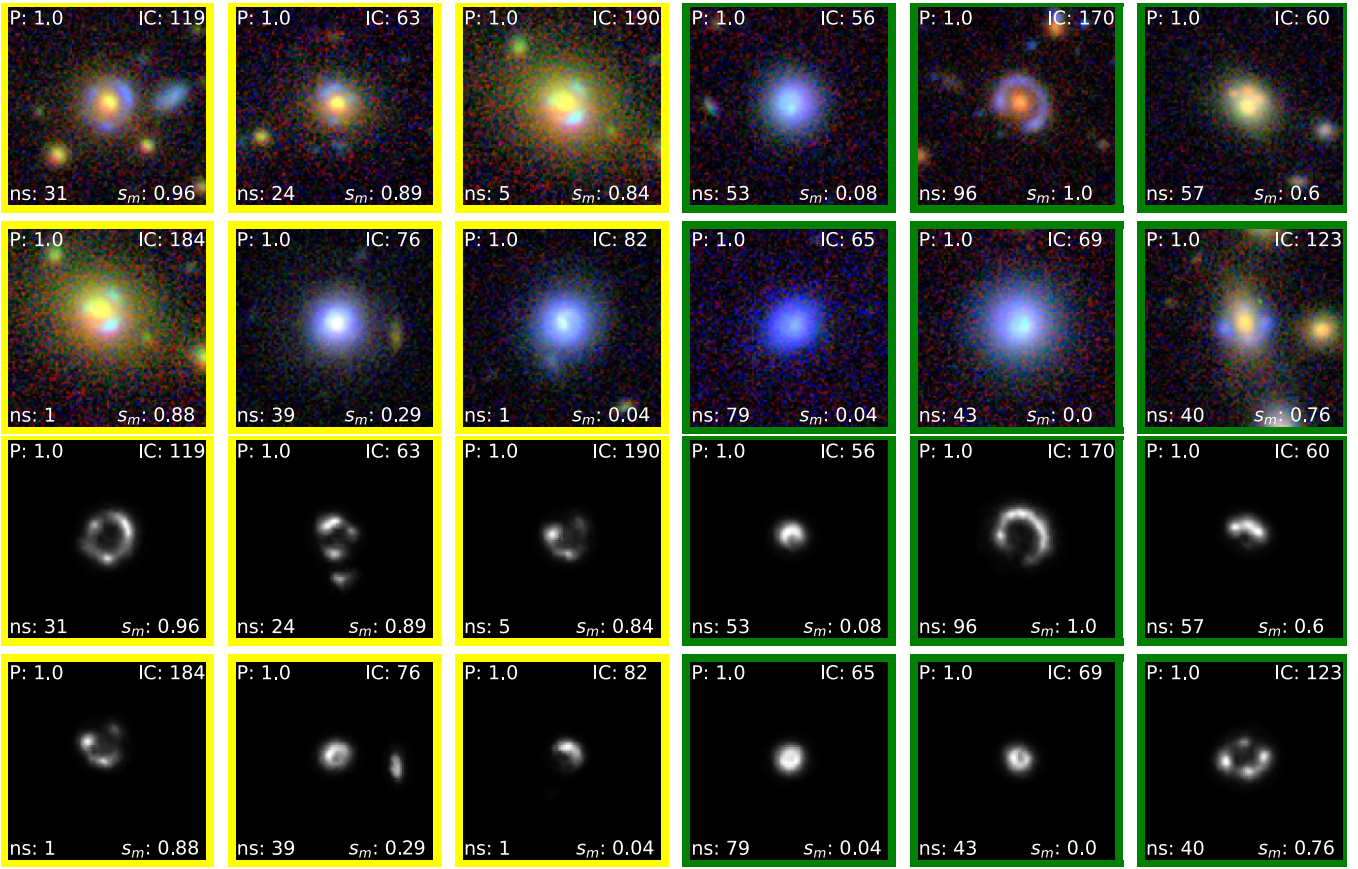

**Figure 10.** Top 2 rows: Top 12 candidates from BG sample rank-ordered based on *P*. Bottom 2 rows: Segmentation maps of the corresponding top 12 candidates. We have shown the candidates with $n_s$ scores $\geq 40$ with the border green and the candidates between $n_s$ scores $> 0$ and $n_s$ scores $<40$ with the image border yellow. The details are shown in Table D6.

**Table 3.** Distribution of TP and FP in the LRG sample with respect to the mean of human classifier votes $s_m$.

| Condition | Red | Yellow | Green | Blue | Total |
| --- | --- | --- | --- | --- | --- |
| $s_m > 0.5$ (TP) | 3 | 13 | 2 | 5 | 23 |
| $s_m \leq 0.5$ (FP) | 53 | 103 | 7 | 14 | 177 |
| Total | 56 | 116 | 9 | 19 | 200 |

Although the decision tree was tailored for the BG we apply the decision tree to an LRG sample, obtaining similar results.

Whereas segmentation can help select genuine lenses, subtracting foreground lens light also significantly aids lens modeling (Nightingale, Dye & Massey 2018; Etherington et al. 2022). Previous studies, such as by 'Pearson, Li & Dye (2019)', have shown a 34 per cent average increase in accuracy of predicted lens model parameters by removing foreground lens light.

## 6 CONCLUSION

In this work, we have introduced a segmentation algorithm (U-Net) to aid in reducing false positives when searching for galaxy–scale strong lenses in large surveys. We add this U-net algorithm to our previous classifier neural network (Nagam et al. 2023) which primarily used *P* and IC to detect and rank-order strong lenses. We illustrate its effectiveness by applying it to a sample of galaxies from the Kilo-Degree Survey.

We generate a mock data-set (Petrillo et al. 2017) of 10 000 mock lens and 10 000 non-lens instances and we applied a classification threshold ($P_{thres} > 0.3$), resulting in the identification of 9586 mock lens candidates and 875 non-lens candidates. Analysing the distribution of candidates, especially the impact of the 'Red' candidates ($n_s = 0$), which could eliminate 77 per cent of false positives at the cost of discarding 14 per cent of true lenses. This highlights the importance of segmentation results, to significantly increase the purity of the final sample of strong lenses.

The final decision tree for the selection of strong gravitational lenses, for the Bright Galaxy (BG) sample from KiDS, involves rank-ordering candidates based on their classification scores (*P*), filtering candidates with Information Content (IC) less than or equal to 50, and removing candidates with zero classified source pixels ($n_s$) and IC values less than or equal to 104. The subsequent human classifier validation process further refines the selection, revealing that eliminating $n_s = 0$ candidates can significantly reduce false positives by only losing considerably fewer confirmable strong lensing candidates. We present fourteen new strong lensing candidates which were discovered with U-Denselens and validated by four or more human classifiers as lens or may-be lens. The extension of the classifier to the Luminous Red Galaxy (LRG) sample confirms the decision tree's effectiveness, demonstrating a reduction in false positives by a quarter. The incorporation of human classifiers in the validation process ensures the preservation of genuine candidates while enhancing the reliability of the selection.

Looking ahead, our study suggests potential avenues for improvement, such as enhancing the realism of training data, incorporating additional lensing types, and expanding the negative data base. Although the classifier is fine-tuned for KiDS *r*-band data, we expect the proposed decision tree to be a robust framework for automation of finding strong gravitational lenses in the upcoming large-scale astronomical surveys e.g, those carried out with Euclid.

**ACKNOWLEDGEMENTS**

We would like to thank the Center for Information Technology of the University of Groningen for their support and for providing access to the Peregrine high performance computing cluster. The research for this paper was funded by the Centre for Data Science and Systems Complexity at the University of Groningen (www.rug.nl/research/fse/themes/dssc/). We would also thank the colloborators Yue-Dong and Rui Li for creating a website to vote for the lensing candidates. CT and VB acknowledge the INAF grant 2022 LEMON.

**DATA AVAILABILITY**

The data used in the paper is available on request. The data underlying this article will be shared on reasonable request to the corresponding author.

**REFERENCES**

Agnello A. et al., 2015, MNRAS, 454, 1260
Akeret J., Chang C., Lucchi A., Refregier A., 2017, Astron. Comput., 18, 35
Akhazhanov A. et al., 2022, MNRAS, 513, 2407
Anguita T. et al., 2018, MNRAS, 480, 5017
Badrinarayanan V., Kendall A., Cipolla R., 2017, IEEE Transact. Pattern Anal. Mach. Intell., 39, 2481
Barbera F. L., de Carvalho R. R., Kohl-Moreira J. L., Gal R. R., Soares-Santos M., Capaccioli M., Santos R., Sant'Anna N., 2008, PASP, 120, 681
Barnacka A., 2018, Phys. Rep., 778–779, 1
Bekki K., 2021, A&A, 647, A120
Belokurov V., Evans N., Hewett P., Moiseev A., McMahon R., Sanchez S., King L., 2009, MNRAS, 392, 104
Bertin E., Arnouts S., 1996, A&AS, 117, 393
Biesiada M., 2006, Phys. Rev. D, 73, 023006
Bolton A. S., Burles S., Koopmans L. V. E., Treu T., Moustakas L. A., 2006, ApJ, 638, 703
Bolton A. S., Burles S., Koopmans L. V. E., Treu T., Gavazzi R., Moustakas L. A., Wayth R., Schlegel D. J., 2008, ApJ, 682, 964
Boucaud A. et al., 2019, MNRAS, 491, 2481
Breiman L., 2001, Mach. Learn., 45, 5
Browne I. et al., 2003, MNRAS, 341, 13
Burke C. J., Aleo P. D., Chen Y.-C., Liu X., Peterson J. R., Sembroski G. H., Lin J. Y.-Y., 2019, MNRAS, 490, 3952
Cabanac R. A. et al., 2007, A&A, 461, 813
Cañameras R. et al., 2021, A&A, 653, L6
Cañameras R. et al., 2020, A&A, 644, A163
Cao Z., Yi Z., Pan J., Su H., Bu Y., Kong X., Luo A., 2023, AJ, 165, 184
Capaccioli M., Schipani P., 2011, The Messenger, 146, 27
Chae K.-H. et al., 2002, Phys. Rev. Lett., 89, 151301
Chae K.-H., Chen G., Ratra B., Lee D.-W., 2004, ApJ, 607, L71
Chan J. H. H., Suyu S. H., Chiueh T., More A., Marshall P. J., Coupon J., Oguri M., Price P., 2015, ApJ, 807, 138
Chan J. H. H. et al., 2016, ApJ, 832, 135
Christ C., Nord B., Gozman K., Ottenbreit K., 2020, in Abstracts of the 235th AAS Meeting. AAS, Honolulu, HI, p. 469
Collett T. E., 2015, ApJ, 811, 20
Davies A., 2022, PhD thesis, The Open University
Davies A., Serjeant S., Bromley J. M., 2019, MNRAS, 487, 5263
de Jong J. T., Verdoes Kleijn G. A., Kuijken K. H., Valentijn E. A., 2013, Exp. Astron., 35, 25
Dewdney P. E., Hall P. J., Schilizzi R. T., Lazio T. J. L. W., 2009, Proc. IEEE, 97, 1482
Diehl H. T. et al., 2017, ApJS, 232, 15
Ellis R. S., 2010, Philo. Transact. R. Soc. A, 368, 967
Etherington A. et al., 2022, MNRAS, 517, 3275
Farias H., Ortiz D., Damke G., Jaque Arancibia M., Solar M., 2020, Astron. Comput., 33, 100420
Gavazzi R., Treu T., Rhodes J. D., Koopmans L. V. E., Bolton A. S., Burles S., Massey R. J., Moustakas L. A., 2007, ApJ, 667, 176
Gavazzi R., Treu T., Koopmans L. V. E., Bolton A. S., Moustakas L. A., Burles S., Marshall P. J., 2008, ApJ, 677, 1046
Gavazzi R., Marshall P. J., Treu T., Sonnenfeld A., 2014, ApJ, 785, 144
Gentile F. et al., 2021, MNRAS, 510, 500
Gilman D., Du X., Benson A., Birrer S., Nierenberg A., Treu T., 2019, MNRAS, 492, L12
Gini C., 1921, Econ. J., 31, 124
Giusarma E., Hurtado M. R., Villaescusa-Navarro F., He S., Ho S., Hahn C., 2019, preprint (arXiv:1910.04255)
Grillo C. et al., 2018, ApJ, 860, 94
Gu M., Wang F., Hu T., Yu S., 2023, in 2023 4th International Conference on Computer Engineering and Application (ICCEA). IEEE, Hangzhou, China, p. 512
Guo D., Qiu B., Liu Y., Xiang G., 2021, in 2021 6th International Conference on Intelligent Computing and Signal Processing (ICSP). IEEE, Xi'an, China, p. 985
He K., Gkioxari G., Dollár P., Girshick R. B., 2017, IEEE, p. 2961, preprint (arXiv:1703.06870)
He Q. et al., 2020a, MNRAS, 496, 4717
He Z. et al., 2020b, MNRAS, 497, 556
Heymans C. et al., 2012, MNRAS, 427, 146
Hezaveh Y. et al., 2013, ApJ, 767, 132
Huang X. et al., 2020, ApJ, 894, 78
Ibata R. A. et al., 2017, ApJ, 848, 128
Inada N., Oguri M., Rusu C. E., Kayo I., Morokuma T., 2014, AJ, 147, 153
Jacobs C., Glazebrook K., Collett T., More A., McCarthy C., 2017, MNRAS, 471, 167
Jaelani A. T. et al., 2020, MNRAS, 494, 3156
Jia P., Liu Q., Sun Y., 2020, AJ, 159, 212
Kochanek C. S. P. L. S., 2003, Carnegie Observat. Astrophys. Ser., 2, 211
Koopmans L. V. E., 2004, preprint (astro-ph/0412596)
Koopmans L., Browne I., Jackson N., 2004, New Astron. Rev., 48, 1085
Koopmans L. V. E. et al., 2009, ApJ, 703, L51
Kuijken K. et al., 2011, The Messenger, 146
Kuijken K. et al., 2019, A&A, 625, A2
Lanusse F., Ma Q., Li N., Collett T. E., Li C.-L., Ravanbakhsh S., Mandelbaum R., Póczos B., 2018, MNRAS, 473, 3895
Laureijs R. J., Duvet L., Sanz I. E., Gondoin P., Lumb D. H., Oosterbroek T., Criado G. S., 2010, in Jr J. M. O., Clampin M. C., MacEwen H. A.eds, Proc. SPIE Conf. Ser. Vol. 7731, Space Telescopes and Instrumentation 2010: Optical, Infrared, and Millimeter Wave. SPIE, Bellingham, p. 453
Lecun Y., Bottou L., Bengio Y., Haffner P., 1998, Proc. IEEE, 86, 2278
Lemon C. et al., 2020, MNRAS, 494, 3491
Li R., Shu Y., Wang J., 2018, MNRAS, 480, 431
Li R. et al., 2020, ApJ, 899, 30
Li R. et al., 2021, ApJ, 923, 16
Long M., Yang Z., Xiao J., Yu C., Zhang B., 2019, in Proc. ASP Conf. Ser. Vol. 523, Astronomical Data Analysis Software and Systems XXVII. Astron. Soc. Pac., San Francisco, p. 123
Meneghetti M., Bartelmann M., Dolag K., Moscardini L., Perrotta F., Baccigalupi C., Tormen G., 2005, A&A, 442, 413
Metcalf R. B. et al., 2019, A&A, 625, A119

Mitchell J. L., Keeton C. R., Frieman J. A., Sheth R. K., 2005, ApJ, 622, 81

Miyazaki S. et al., 2012, in McLean I. S., Ramsay S. K., Takami H.eds, Proc. SPIE Conf. Ser. Vol. 327, Ground-based and Airborne Instrumentation for Astronomy IV. SPIE, Bellingham, p. 327

More A., Cabanac R., More S., Alard C., Limousin M., Kneib J.-P., Gavazzi R., Motta V., 2012, ApJ, 749, 38

More A. et al., 2015, MNRAS, 455, 1191

More A. et al., 2016, MNRAS, 465, 2411

Myers S. et al., 1999, AJ, 117, 2565

Myers S. T. et al., 2001, Astronomical society of the Pacific, 237, 51

Nadler E. O., Birrer S., Gilman D., Wechsler R. H., Du X., Benson A., Nierenberg A. M., Treu T., 2021, ApJ, 917, 7

Nagam B. C., Koopmans L. V. E., Valentijn E. A., Kleijn G. V., de Jong J. T. A., Napolitano N., Li R., Tortora C., 2023, MNRAS, 523, 4188

Narayan S., 1997, Inform. Sci., 99, 69

Negrello M. et al., 2010, Science, 330, 800

Nightingale J. W., Dye S., Massey R. J., 2018, MNRAS, 478, 4738

Nightingale J. W., Massey R. J., Harvey D. R., Cooper A. P., Etherington A., Tam S.-I., Hayes R. G., 2019, MNRAS, 489, 2049

Nord B., Buckley-Geer E. J., Lin H., Diehl H. T., Gaitsch H., 2015, 225, 255.21

Nord B. et al., 2016, ApJ, 827, 51

Nord B. et al., 2020, MNRAS, 494, 1308

Oguri M. et al., 2005, ApJ, 622, 106

Oguri M. et al., 2006, AJ, 132, 999

Oguri M. et al., 2008a, AJ, 135, 512

Oguri M. et al., 2008b, AJ, 135, 520

Ostdiek B., Diaz Rivero A., Dvorkin C., 2022a, A&A, 657, L14

Ostdiek B., Diaz Rivero A., Dvorkin C., 2022b, ApJ, 927, 83

Pawase R. S., Courbin F., Faure C., Kokotanekova R., Meylan G., 2014, MNRAS, 439, 3392

Pearson J., Pennock C., Robinson T., 2018, Emergent Sci., 2, 1

Pearson J., Li N., Dye S., 2019, MNRAS, 488, 991

Petrillo C. E. et al., 2017, MNRAS, 472, 1129

Petrillo C. E. et al., 2019a, MNRAS, 482, 807

Petrillo C. E. et al., 2019b, MNRAS, 484, 3879

Pourrahmani M., Nayyeri H., Cooray A., 2018, ApJ, 856, 68

Qi J., Chen M., Wu Z., Su C., Yao X., Mei C., 2022, in China Automation Congress (CAC). IEEE, p. 1901

Quinn P., Axelrod T., Bird I., Dodson R., Szalay A., Wicenec A., 2015, preprint (arXiv:1501.05367)

Ren S., He K., Girshick R., Sun J., 2015, in Cortes C., Lawrence N., Lee D., Sugiyama M., Garnett R., eds, Advances in Neural Information Processing Systems. Curran Associates, Inc., NY, United States, p. 91, https://proceedings.neurips.cc/paper_files/paper/2015/file/14bfa6bb14875e45bba028a21ed38046-Paper.pdf

Rezaei S., McKean J., Biehl M., de W., Lafontaine A., 2022, MNRAS, 517, 1156

Rhee G., 1991, Nature, 350, 211

Riggi S. et al., 2023, Astron. Comput., 42, 100682

Rojas K. et al., 2022, A&A, 668, A73

Ronneberger O., Fischer P., Brox T., 2015, Springer International Publishing, part III 18 , p. 234

Sarbu N., Rusin D., Ma C.-P., 2001, ApJ, 561, L147

Savary E. et al., 2021

Savary E. et al., 2022, A&A, 666, A1

Schaefer C., Geiger M., Kuntzer T., Kneib J.-P., 2018, A&A, 611, A2

Şengül A., Dvorkin C., 2022, MNRAS, 516, 336

Sereno M., 2002, A&A, 393, 757

Serjeant S., 2014, ApJ, 793, L10

Shu Y., Cañameras R., Schuldt S., Suyu S. H., Taubenberger S., Inoue K. T., Jaelani A. T., 2022, A&A, 662, A4

Shu Y., Bolton A. S., Brownstein J. R., 2015, ApJ, 803, 2

Shu Y., Brownstein J. R., Bolton A. S., Koopmans L. V. E., 2017, ApJ, 851, 48

Simpson R., Page K. R., De Roure D., 2014, in Proceedings of the 23rd International Conference on World Wide Web. ACMDL, p. 1049

Spiniello C., Barnabé M., Koopmans L. V. E., Trager S. C., 2015, MNRAS, 452, L21

Sygnet J. F., Tu H., Fort B., Gavazzi R., 2010, A&A, 517, A25

Tanaka M. et al., 2016, ApJ, 826, L19

Tanoglidis D. et al., 2022, Astro. Comput., 39, 100580

Dark Energy Survey Collaboration, 2005, preprint(astro-ph/0510346)

Tortora C., Napolitano N. R., Romanowsky A. J., Jetzer P., 2010, ApJ, 721, L1

Treu T., Ellis R. S., 2015, Contemp. Phys., 56, 17

Treu T., Koopmans L. V. E., 2002, ApJ, 575, 87

Treu T., Shajib A. J., 2023, Springer Nature, Singapore, p. 251

Treu T. et al., 2018, MNRAS, 481, 1041

Treu T., Suyu S. H., Marshall P. J., 2022, A&AR, 30, 8

Turner E. L., Ostriker J. P., Gott III J. R., 1984, ApJ, 284, 1

Turyshev S. G., Toth V. T., 2022, MNRAS, 513, 5355

Tyson J. A., 2002, in Tyson J. A., Wolff S.eds, Proc. SPIE Conf. Ser. Vol. 4836, Survey and Other Telescope Technologies and Discoveries. SPIE, Bellingham, p. 10

Vojtekova A., Lieu M., Valtchanov I., Altieri B., Old L., Chen Q., Hroch F., 2020, MNRAS, 503, 3204

Wang Y. et al., 2022, ApJ, 928, 1

Wardlow J. L. et al., 2012, ApJ, 762, 59

Weiner C., Serjeant S., Sedgwick C., 2020, Res. Notes Am. Astron. Soc., 4, 190

Widyaningrum R., Candradewi I., Aji N. R. A. S., Aulianisa R., 2022, Imaging Sci. Dent., 52, 383

Wong K., HSC SSP Strong Lens Working Group, 2018, in American Astronomical Society Meeting Abstracts #231. AAS, p. 226.01

Wootten A., 2003, in Large Ground-Based Telescopes. SPIE, Vol. 4837, p. 110

Wu T., 2020, in 2020 International Conference on Big Data Artificial Intelligence Software Engineering (ICBASE). IEEE, p. 390

Wu C. et al., 2018, MNRAS, 482, 1211

Zhan H., 2018, 42nd COSPAR Scientific Assembly, 42, E1

Zhang T.-J., 2004, ApJ, 602, L5

Zitrin A. et al., 2012, ApJ, 749, 97

## APPENDIX A: SEGMENTATION TRAINING

The U-Net algorithm is trained for positive candidates (lenses) with source pixels as 1 and the rest of the pixels as 0. This is explained in Fig. A1 (top). We generate the mock lens by adding mock lensed source on top of LRG (see appendix A from Nagam et al. 2023) and hence the output segmentation map for each training sample is created by designating all the pixels belonging to the lensed source as 1 and rest to be zero. For non-lenses, all the pixels are trained with label 0 as shown in Fig. A1 (bottom). Our training data-set comprised of 100 000 mock strong lenses and ∼6000 non-lenses (refer Nagam et al. 2023) for training, testing and validation. We have trained a single U-Net network (not an ensemble) for about 2500 iterations. Our batch size for each training comprised of 64 samples for training and 32 samples for validation.

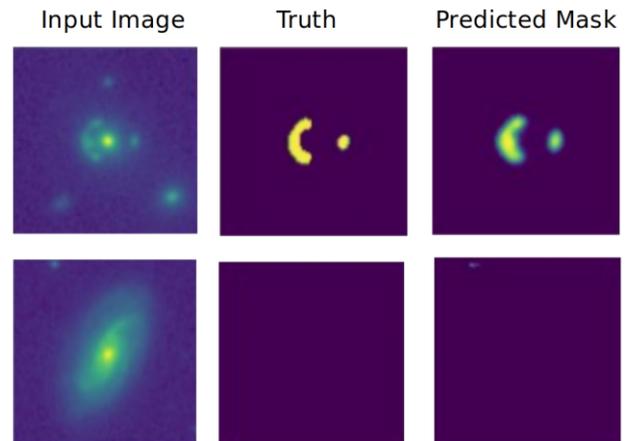

**Figure A1.** The training of lenses and non-lenses for the U-Net algorithm. For training, the source pixels are labelled as 1 and rest of the pixels are labelled as 0 (top). For non-lenses, all the pixels are labelled as 0 for the training (below).

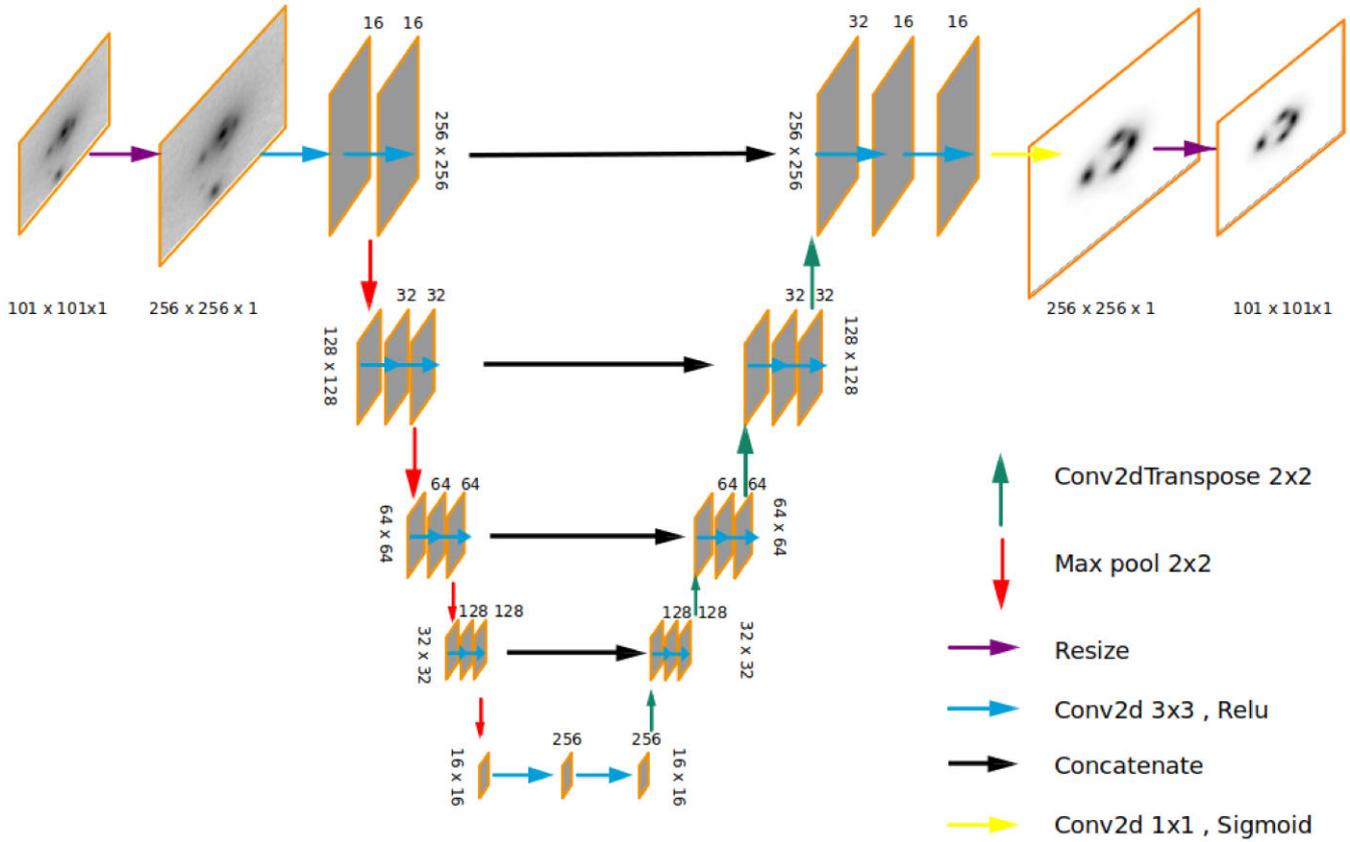

**Figure A2.** Description of UNet architecture used in this paper. Thus the final output obtained from the U-net model has $101 \times 101$ pixels with each pixel varying between 0 and 1. In our model, we resize our input image of $101 \times 101$ pixels into the shape of $256 \times 256$ pixels through interpolation. Then the image is passed through the U-Net architecture.

## APPENDIX B: MOCK LENSES

We have shown a sample of mock lenses used in the research. Fig. 8 (top) and (bottom) shows the top 25 rank-ordered candidates and its respective segmentation maps from the mock data.

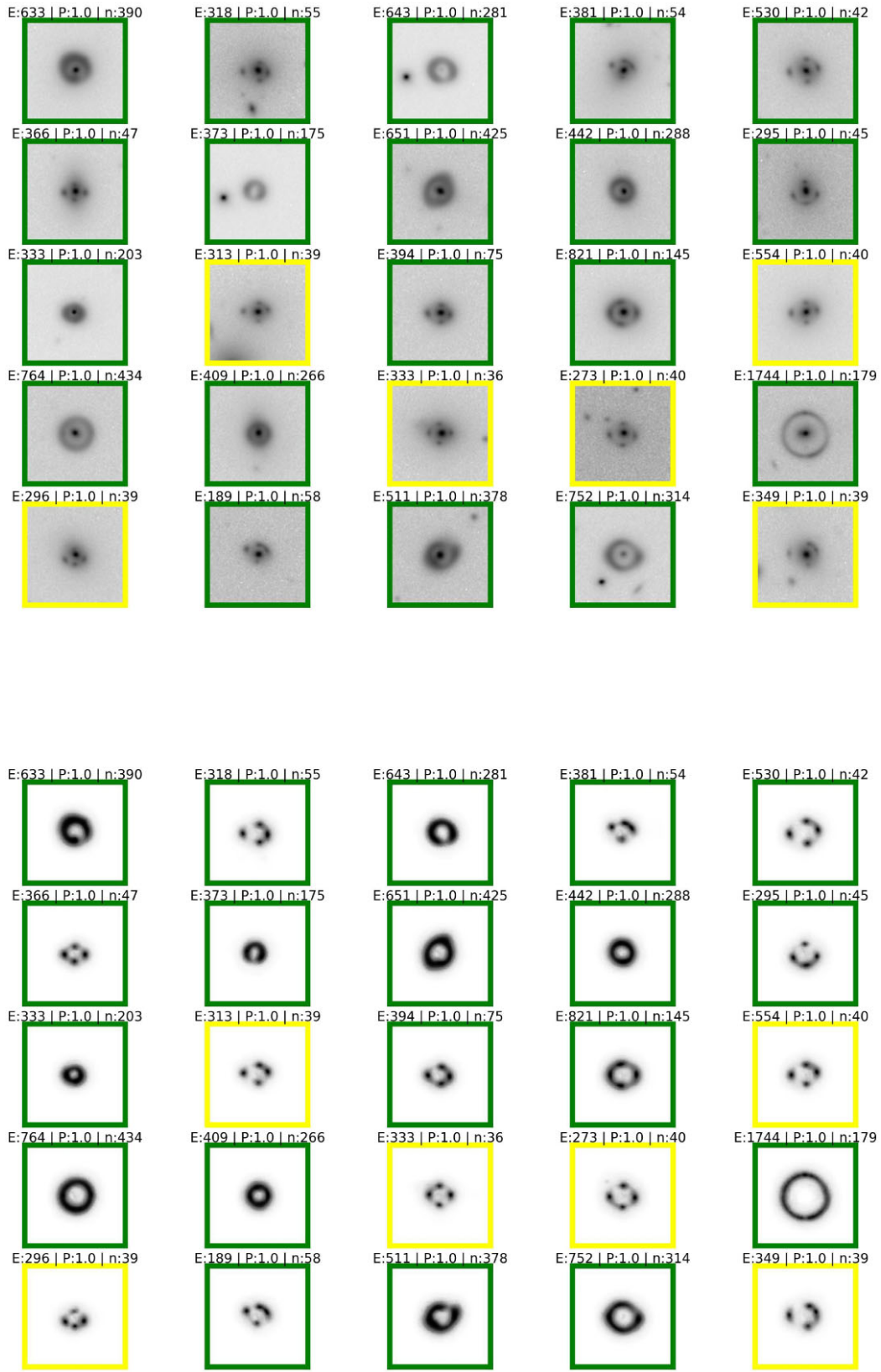

**Figure B1.** Illustration of DenseLens results for mock data. **Top:** Classification prediction scores ($P$), IC ($E$), total number of classified source pixels ($n_s$) for first 25 candidates ran-ordered by $P$ value. **Bottom:** The bottom plot shows the segmentation results for the respective candidates shown above.

## APPENDIX C: OTHER APPROACHES

We conducted an experiment involving a four-CNN network ensemble, utilizing segmentation maps instead of traditional images, to assess the potential performance improvements with this alternative input. Surprisingly, our findings revealed a decrease in performance when the networks were trained with segmentation maps. This leads us to the conclusion that CNNs trained on segmentation maps may struggle due to the inherently reduced information content compared to original cutout images.

It is evident that the CNNs trained on original cutouts exhibit superior learning capabilities, capturing additional crucial information such as lens properties, background objects within the field, and their respective intensities. This underscores the importance of utilizing original cutout images for training, as they provide a richer data-set for the networks to learn and comprehend the complexities of the visual data at hand.

## APPENDIX D: TABLES

In this section, we primarily show the KiDS tile name, RA, Dec., P, IC, and $n_s$ values for the candidates displayed in Sec 4

**Table D1.** KiDS tile name, RA, Dec., P, IC, and $n_s$ values for the candidates with IC ≤ 50 shown in Fig. 2.

| KiDS tile name | RA | Dec. | P | IC | $n_s$ |
| --- | --- | --- | --- | --- | --- |
| KIDS_204.0_0.5 | 203.747796 | 0.271528 | 0.99967 | 43.30 | 44 |
| KIDS_195.0_−0.5 | 194.710143 | −0.224885 | 0.99960 | 41.66 | 45 |
| KIDS_349.6_−30.2 | 350.173666 | −30.600492 | 0.99945 | 43.79 | 24 |
| KIDS_335.5_−31.2 | 335.055485 | −31.259669 | 0.99886 | 44.76 | 0 |
| KIDS_231.0_0.5 | 231.114952 | 0.838905 | 0.99860 | 46.60 | 13 |
| KIDS_33.6_−34.1 | 33.98169 | −34.407147 | 0.99846 | 43.73 | 0 |
| KIDS_25.1_−29.2 | 24.558549 | −29.430491 | 0.99832 | 22.60 | 20 |
| KIDS_353.2_−29.2 | 353.474196 | −29.531741 | 0.99809 | 46.35 | 26 |
| KIDS_46.7_−29.2 | 46.272839 | −29.246367 | 0.99802 | 43.61 | 22 |
| KIDS_24.0_−34.1 | 24.492328 | −33.998216 | 0.99801 | 45.72 | 25 |
| KIDS_222.6_2.5 | 222.580281 | 2.125373 | 0.99797 | 49.78 | 2 |
| KIDS_349.7_−29.2 | 349.463897 | −29.616792 | 0.99783 | 46.88 | 75 |

**Table D2.** KiDS tile name, RA, Dec., P, IC, and $n_s$ values for the top 6 candidates in each of the four regimes (blue, yellow, green, and red) with IC ≤ 50 shown in Fig. 5.

| KiDS tile name tag | RA | Dec. | P | IC | $n_s$ |
| --- | --- | --- | --- | --- | --- |
| KIDS_16.3_−31.2 | 16.770493 | −31.478064 | 0.98284 | 190 | 0 |
| KIDS_130.0_−1.5 | 129.889052 | −1.679115 | 0.96702 | 119 | 0 |
| KIDS_159.4_−2.5 | 159.780754 | −2.275022 | 0.99808 | 137 | 0 |
| KIDS_182.0_−1.5 | 181.93025 | −1.065401 | 0.99826 | 115 | 0 |
| KIDS_11.7_−31.2 | 11.559175 | −31.340386 | 0.97319 | 215 | 0 |
| KIDS_225.0_−1.5 | 225.357979 | −1.491227 | 0.98618 | 154 | 0 |
| KIDS_47.1_−27.2 | 46.665370 | −27.607481 | 0.9999 | 119 | 31 |
| KIDS_205.0_−1.5 | 204.68696 | −1.151899 | 0.957320 | 135 | 21 |
| KIDS_24.2_−30.2 | 23.697086 | −29.947897 | 0.999987 | 63 | 24 |
| KIDS_2.4_−32.1 | 2.066701 | −32.621056 | 0.999813 | 184 | 1 |
| KIDS_134.0_−1.5 | 133.693949 | −1.360288 | 0.975157 | 138 | 7 |
| KIDS_189.0_0.5 | 188.97575 | 0.930670 | 0.999107 | 69 | 6 |
| KIDS_15.6_−34.1 | 15.366003 | −33.722057 | 0.99993 | 170 | 96 |
| KIDS_216.6_−2.5 | 216.732485 | −2.622653 | 0.97750 | 140 | 135 |
| KIDS_45.4_−31.2 | 44.920964 | −31.037683 | 0.996105 | 108 | 85 |
| KIDS_345.6_−34.1 | 345.609561 | −33.943769 | 0.99953 | 123 | 40 |
| KIDS_182.0_−0.5 | 182.322335 | −0.571087 | 0.980950 | 141 | 53 |
| KIDS_3.5_−31.2 | 3.020876 | −30.684705 | 0.99897 | 225 | 82 |
| KIDS_349.3_−33.1 | 349.231535 | −32.986890 | 0.988414 | 51 | 0 |
| KIDS_344.9_−31.2 | 345.395628 | −31.543209 | 0.99609 | 74 | 0 |
| KIDS_189.0_0.5 | 188.614697 | 0.910346 | 0.992001 | 83 | 0 |
| KIDS_12.0_−34.1 | 12.573356 | −34.283909 | 0.962898 | 83 | 0 |
| KIDS_31.2_−34.1 | 31.476136 | −34.338695 | 0.988035 | 54 | 0 |
| KIDS_2.4_−34.1 | 2.084894 | −34.455275 | 0.98812 | 55 | 0 |

**Table D3.** KiDS tile name, RA, Dec., P, IC, $n_s$, and tag values for the LRG candidates shown in Fig. 7.

| KiDS tile name | RA | Dec. | P | IC | $n_s$ | tag |
| --- | --- | --- | --- | --- | --- | --- |
| KIDS_130.0_−1.5 | 129.889052 | −1.679115 | 0.967 | 119 | 0 | b1 |
| KIDS_333.9_−33.1 | 333.412536 | −33.198985 | 0.945 | 130 | 0 | b2 |
| KIDS_159.4_−2.5 | 159.780754 | −2.275022 | 0.998 | 137 | 0 | b3 |
| KIDS_31.2_−34.1 | 31.619539 | −33.796853 | 0.948 | 122 | 0 | b4 |
| KIDS_182.0_−1.5 | 181.93025 | −1.065401 | 0.998 | 115 | 0 | b5 |
| KIDS_336.3_−28.2 | 336.083708 | −27.841762 | 0.968 | 102 | 0 | b6 |
| KIDS_47.1_−27.2 | 46.66537 | −27.607481 | 1 | 119 | 31 | y1 |
| KIDS_189.0_0.5 | 188.97575 | 0.93067 | 0.999 | 69.3 | 6 | y2 |
| KIDS_134.0_−1.5 | 133.693949 | −1.360288 | 0.975 | 138 | 7 | y3 |
| KIDS_168.0_0.5 | 168.224904 | 0.179072 | 0.999 | 87.4 | 28 | y4 |
| KIDS_205.0_−1.5 | 204.686617 | −1.151341 | 0.998 | 149 | 22 | y5 |
| KIDS_24.2_−30.2 | 23.607086 | −29.947897 | 1 | 62.8 | 24 | y6 |
| KIDS_185.5_2.5 | 185.672168 | 2.099758 | 0.993 | 154 | 40 | g1 |
| KIDS_181.5_−2.5 | 181.196706 | −2.720097 | 0.998 | 385 | 84 | g2 |
| KIDS_0.0_−35.1 | 0.292592 | −35.259432 | 0.991 | 80.2 | 42 | g3 |
| KIDS_215.0_−0.5 | 214.827121 | −0.420325 | 1 | 60.0 | 57 | g4 |
| KIDS_184.0_−0.5 | 183.600158 | −0.535089 | 0.980 | 81.5 | 0 | r1 |
| KIDS_31.2_−34.1 | 31.476136 | −34.338695 | 0.988 | 53.6 | 0 | r2 |
| KIDS_214.6_2.5 | 214.367082 | 1.993100 | 0.998 | 52.3 | 0 | r3 |

**Table D4.** KiDS tile name, RA, Dec., P, IC, $n_s$, $s_m$, and $s_{std}$ values for high std. dev. candidates of LRG sample shown in Fig. 8.

| KiDS tile name | RA | Dec. | P | IC | $n_s$ | $s_m$ | $s_{std}$ |
|---|---|---|---|---|---|---|---|
| KIDS_238.0_−1.5 | 238.000448 | −1.785379 | 0.994 | 79.8 | 15 | 0.5 | 0.4071 |
| KIDS_184.0_−0.5 | 183.600158 | −0.535089 | 0.98 | 81.5 | 0 | 0.5375 | 0.3662 |
| KIDS_48.9_−31.2 | 48.819245 | −31.131718 | 0.993 | 65.9 | 3 | 0.4375 | 0.3623 |
| KIDS_34.5_−33.1 | 34.78658 | −33.52217 | 0.985 | 61.3 | 15 | 0.4375 | 0.3623 |
| KIDS_1.2_−34.1 | 1.569209 | −34.476066 | 0.984 | 94.6 | 4 | 0.2625 | 0.3623 |
| KIDS_185.0_0.5 | 185.354352 | 0.964915 | 0.947 | 253 | 66 | 0.325 | 0.3615 |

**Table D5.** KiDS tile name, RA, Dec., P, IC, $n_s$, and $s_m$, values for fourteen newly found strong lensing candidates in the BG data shown in Fig. 9.

| KiDS tile name | RA | DEC | P | IC | $n_s$ | $s_m$ |
|---|---|---|---|---|---|---|
| KIDS_32.2_−30.2 | 32.135 | −30.5285 | 1 | 58 | 41 | 0.5 |
| KIDS_45.4_−31.2 | 44.921 | −31.0377 | 0.996 | 108 | 85 | 0.76 |
| KIDS_19.6_−30.2 | 19.3416 | −30.0427 | 0.99 | 155 | 127 | 0.66 |
| KIDS_177.0_−0.5 | 177.209 | −0.19919 | 0.99 | 181 | 50 | 0.64 |
| KIDS_187.0_−1.5 | 187.342 | −1.0351 | 0.98 | 221 | 66 | 0.64 |
| KIDS_216.6_−2.5 | 216.732 | −2.62265 | 0.98 | 140 | 135 | 0.78 |
| KIDS_2.4_−32.1 | 2.067021 | −32.620669 | 1 | 190 | 5 | 0.84 |
| KIDS_215.6_−2.5 | 215.648 | −2.64927 | 0.99 | 58 | 8 | 0.78 |
| KIDS_44.9_−30.2 | 44.9496 | −30.4085 | 0.98 | 135 | 7 | 0.76 |
| KIDS_40.4_−33.1 | 40.0624 | −32.9086 | 0.98 | 61 | 13 | 0.65 |
| KIDS_2.4_−33.1 | 2.41595 | −33.1302 | 0.96 | 84 | 8 | 0.69 |
| KIDS_4.8_−34.1 | 5.31821 | −33.7695 | 0.96 | 130 | 20 | 0.63 |
| KIDS_333.8_−29.2 | 333.357 | −29.3585 | 0.96 | 58 | 2 | 0.69 |
| KIDS_208.0_−0.5 | 207.860973 | −0.199983 | 0.96 | 72 | 9 | 0.51 |

**Table D6.** KiDS tile name, RA, Dec., P, IC, $n_s$, and $s_m$ values for top 12 candidates from BG sample rank-ordered based on P shown in Fig. 10.

| KiDS tile name | RA | Dec. | P | IC | $n_s$ | $s_m$ |
|---|---|---|---|---|---|---|
| KIDS_47.1_−27.2 | 46.66537 | −27.607481 | 0.99999 | 119 | 31 | 0.96 |
| KIDS_24.2_−30.2 | 23.607086 | −29.947897 | 0.99999 | 63 | 24 | 0.89 |
| KIDS_2.4_−32.1 | 2.067021 | −32.620669 | 0.99999 | 190 | 5 | 0.84 |
| KIDS_164.0_−0.5 | 163.792396 | −0.68975 | 0.99994 | 56 | 53 | 0.08 |
| KIDS_15.6_−34.1 | 15.366003 | −33.722057 | 0.99993 | 170 | 96 | 1.0 |
| KIDS_215.0_−0.5 | 214.827121 | −0.420325 | 0.99988 | 60 | 57 | 0.6 |
| KIDS_2.4_−32.1 | 2.066701 | −32.621056 | 0.99981 | 184 | 1 | 0.88 |
| KIDS_30.6_−32.1 | 31.079838 | −32.007072 | 0.99976 | 76 | 39 | 0.29 |
| KIDS_49.2_−34.1 | 48.764524 | −34.404492 | 0.99973 | 82 | 1 | 0.04 |
| KIDS_225.0_0.5 | 224.895194 | 0.407729 | 0.99973 | 65 | 79 | 0.04 |
| KIDS_42.8_−33.1 | 42.709956 | −33.576519 | 0.99956 | 69 | 43 | 0 |
| KIDS_345.6_−34.1 | 345.609561 | −33.943769 | 0.99953 | 123 | 40 | 0.76 |

## APPENDIX E: COMPUTATION OF FEATURE IMPORTANCE OF RANDOM FORESTS

Feature importance for Random Forests (Breiman 2001) is calculated based on the following steps. The Gini impurity (Gini 1921) measures the probability of incorrectly classifying a random chosen element in the data-set and the impurity reduction is computed as the weighted difference between the impurity of the current node and the impurity of the child nodes after a split.

$$\Delta I_{Gini} = I_{Gini}(\text{parent}) - \left( \frac{N_{\text{left}}}{N_{\text{parent}}} \cdot I_{Gini}(\text{left}) + \frac{N_{\text{right}}}{N_{\text{parent}}} \cdot I_{Gini}(\text{right}) \right), \quad \text{(E1)}$$

where,

$I_{Gini}(\text{parent})$ is the Gini impurity of the parent node,
$N_{\text{left}}$ and $N_{\text{right}}$ are the number of samples in the left and right child nodes, respectively,
$I_{Gini}(\text{left})$ and $I_{Gini}(\text{right})$ are the Gini impurities of the left and right child nodes, respectively.

For each tree in the ensemble, impurity reduction for each individual feature (f) across all split nodes of the tree are summed.

$$\text{Importance}_{\text{tree, feature}} = \sum_{\text{nodes}} \text{Impurity Reduction}_{\text{node, feature}}.$$

Sum up the importance scores across all trees for each feature obtained from each tree in the ensemble.

$$\text{Importance}_{\text{feature}} = \frac{1}{N} \sum_{\text{trees}} \text{Importance}_{\text{tree, feature}}$$

Then a normalized importance for each feature is calculated

$$\text{Norm. Importance}_{\text{feature}} = \frac{\text{Importance}_{\text{feature}}}{\sum_{\text{features}} \text{Importance}_{\text{feature}}}$$

This paper has been typeset from a T<sub>E</sub>X/LAT<sub>E</sub>X file prepared by the author.